\documentclass[a4paper]{article}
\usepackage{times}
\usepackage{graphicx}
\usepackage{amsmath, amsfonts,amssymb}
\usepackage{setspace}
\begin{document}

\renewcommand*{\thefootnote}{\arabic{footnote}}
\setcounter{footnote}{0}
\renewcommand*{\thefootnote}{\fnsymbol{footnote}}
\centerline{\Large\bf On the depth of cylindrical indentation of an elastic half-space}
\centerline{\Large\bf for two types of displacement constraints}
\renewcommand*{\thefootnote}{\arabic{footnote}}
\setcounter{footnote}{0}
\vskip3mm
\centerline{\bf{Vlado A. Lubarda${}^{1^*}$ and Marko V. Lubarda${}^{2}$}}

\centerline{${}^{1}$Department of NanoEngineering, UC San Diego, USA}
\centerline{${}^{2}$Department of Mechanical and Aerospace Engineering, UC San Diego, USA}
\vskip2mm
\centerline{${}^*$Corresponding author (Email: vlubarda@ucsd.edu)}

\numberwithin{equation}{section}

\vskip5mm

\begin{abstract}For cylindrical indentation of elastic half-space the relationship between the depth of indentation $\delta$ and the applied force $F$ is nonlinear, in contrast
to the linear relationship between the height of the contact zone $\delta_0$ and the force $F$. While the latter is independent of the boundary conditions used to specify the rigid-body translation, the former depends on a selected datum for vertical displacement. The depth of the indentation is determined for any permissible value of the length $b$, which specifies the points of the free surface where the vertical displacement is required to be zero, $w(\pm b)=0$. From the condition that the work of the indentation force is equal to the work of the contact pressure, it follows that the indentation is geometrically and physically possible  under imposed  boundary conditions $w(\pm b)=0$ provided that $b\ge b_{\rm min}$. The numerical value of  $b_{\rm min}$ is found to be about 10 times greater than the semi-width of the contact zone $a$, based on the
numerical precision in fulfilling the work condition $W_F=W_p$. If a datum is taken to be at a point at some distance $h$ below the load, there is an alternative closed-form expression for $\delta$ in terms of $F$, which involves the Poisson ratio. For $\nu=1/3$, it is found that $h_{\rm min}$ is about $21a$.
A simple expression relating the permissible values of $h$ and $b$ is derived, which is linear for large values of $h$ and $b$.

\end{abstract}

\section{Introduction}

The depth $\delta$ of spherical and conical indentations of an elastic half-space is uniquely related to the indentation force $F$, because a datum for vertical displacement can be taken to be at infinity, where the displacement goes to zero with the distance $r$ from the load as $1/r$. For spherical indentation $F=(4\sqrt{R}/3)E_*\delta^{3/2}$, where $E_*=E/(1-\nu^2)$ is the effective modulus of elasticity, expressed in terms of Young's modulus of elasticity $E$ and Poisson's ratio $\nu$, and $R$ is the radius of a spherical indenter.
The corresponding height of the contact zone is $\delta_0=\delta/2$. For conical indentation, the quadratic relationship $F=(2\tan\alpha/\pi)E_*\delta^2$ holds, where $\alpha$ is the cone angle, while the corresponding height of the contact zone is $\delta_0=2\delta/\pi$.
In contrast to these well-known results \cite{Gladwell}-\cite{Barber}, the relationship between $F$ and $\delta$ for cylindrical indentation of a half-space has not been elaborated upon in the literature, probably because it was recognized that the displacements logarithmically increase with the distance from the load, preventing a datum to be taken at infinity,
which means that $\delta$ depends on a selected point at which the vertical displacement is required to be zero. Most commonly, this datum is taken to be at some vertical distance $h$ below the load, or through a pair of points on the free surface at some horizontal distance $b$ from the center of the load. For finite-size bodies, the depth of indentation also depends on the size and shape of indented body and the location of the indenter relative to the boundaries of the body \cite{Johnson,Barber}. Indentation of a thin elastic layer bonded to a rigid
substrate has been analyzed in \cite{Meijers,Green}, while the study of the indentation by a rigid circular cylinder of finite length has been reported in \cite{Norden,Popov}.
The surface tension effects were considered in \cite{Jia}-\cite{Li}, the indentation of a functionally graded half-space in \cite{Vasu,Gian}, and the micropolar elasticity effects in \cite{Zisis}.

In this paper we derive the closed-form relationships between the depth of indentation and the applied force for cylindrical indentation of a half-space corresponding to different displacement datums and show that these relationships are nonlinear. The datum is first taken to be at the points $x=\pm b$ of the free-surface of a half-space.
The requirement that the work done by the indentation force must be equal to the work done by the contact pressure defines the
minimum value of $b$ for which the indentation is geometrically and physically possible. This minimum value is found to be about 10 times greater than the semi-width of the contact zone $a=(4FR/\pi E_*)^{1/2}$. The corresponding minimum value of the indentation depth is $\delta_{\rm min}\approx 3.5\delta_0$, where $\delta_0=2F/\pi E_*$ is the height of the contact zone.
When the datum is taken to be at a point below the load, at some distance $h$ from the free surface, there is an alternative closed-form expression for $\delta$ in terms of $F$,
which involves the Poisson ratio $\nu$. The minimum permissible value of $h$ is $h_{\rm min}\approx 16.5a$ in the case $\nu=0$, and $h_{\rm min}\approx 27a$ in the case $\nu=1/2$,
with the corresponding minimum indentation depth $\delta_{\rm min}\approx 3.5\delta_0$. The relationship between $h$ and $b$, corresponding to all indentation depths $\delta\ge \delta_{\rm min}$, is shown to be linear for large values of $h$ and $b$. The derived results are useful for the analysis of the indentation of large elastic blocks bonded to a rigid substrate either laterally or at its bottom.

\section{Indentation by a rigid circular cylinder}

\begin{figure}[h!]
\centering{
\includegraphics[scale=0.65]{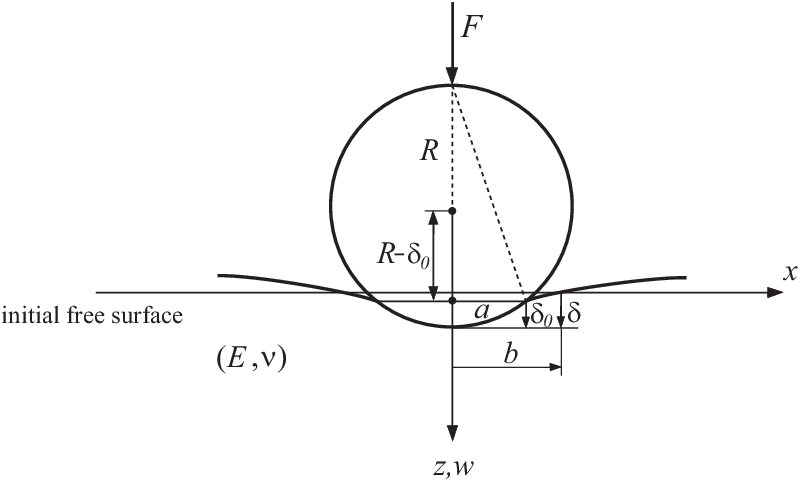}
}
\caption{\label{Fig1} A rigid circular cylinder of radius $R$ pressed into an elastic half-space by a force $F$ (per unit length of the cylinder). The height of the cylindrical cap in contact with the material of a half-space is $\delta_0$, and $2a$ is the corresponding width of the contact. The depth of the indentation relative to the initial level of the free surface ($x$ axis) is $\delta$. The imposed displacement boundary condition is $w(\pm b)=0$, where $b\ge b_{\rm min}$ and $b_{\rm min}$ is determined in the body of the paper. The elastic constants of a half-space are $E$ and $\nu$.
}
\end{figure}

Figure \ref{Fig1} shows the cross section of a smooth rigid cylindrical indenter of radius $R$ pressed against an isotropic elastic half-space whose elastic constants are $E$ and $\nu$. The depth of the indentation due to the vertical force $F$ (per unit length of an infinitely long cylinder), relative to the initial level of the free surface ($x$ axis), is denoted by $\delta$. The value of $\delta$ corresponds to a selected value of $b\ge b_{\rm min}$, which defines the points $x=\pm b$ of the free surface at which the vertical displacement is required to be zero, $w(x=\pm b)=0$. The height of the cylindrical cap in contact with the material of a half-space is $\delta_0$ (indenter penetration depth into the material), and $2a$ is the corresponding width of the contact. The objective is to determine $b_{\rm min}$ and $\delta_{\rm min}$, and
the relationship $\delta=\delta(F,b)$ for $b\ge b_{\rm min}$. The height of the contact zone $\delta_0$ and the relationship $\delta_0=\delta_0(F)$ are independent of the
vertical rigid-body translation, and are thus independent of $b$ as well.

The equation of the circular contact in the $(x,w)$ coordinate system for shallow indentation is
\begin{equation}\label{2.1}
w(x)=\delta-\delta_0\frac{x^2}{a^2}\,,\quad |x|\le a\,,
\end{equation}
where $w(0)=\delta$ and $w(\pm a)=\delta-\delta_0$ (Fig. \ref{Fig1}).
By basic geometry, the contact semi-width $a$ is the geometric mean of $\delta_0$ and $2R-\delta_0$, which for $\delta_0\ll R$ simplifies to $a^2=2R\delta_0$, independently of $b$.
A cylindrical indenter exerts a pressure $p=p(x)$ over the surface of a half-space within the contact width $2a$, which is statically equivalent to applied force $F$.
In linear elasticity this contact pressure is semi-elliptical in shape and given by \cite{Gladwell}-\cite{Barber}
\begin{equation}\label{2.2}
p(x)=p_0\left(1-\frac{x^2}{a^2}\right)^{1/2}\,,\quad p_0=\frac{4}{\pi}\,\bar{p}\,.
\end{equation}
The height  of the contact zone $\delta_0$ can then be obtained from the displacement expression of the Flamant's problem of a concentrated force
on the boundary of a half-space \cite{Timosh} by using the superposition,
\begin{equation}\label{2.3}
\delta_0=\frac{2}{\pi E_*}\int_0^ap(x)\ln\left(\frac{a^2}{x^2}-1\right)\mathrm{d}x\,.
\end{equation}
The substitution of (\ref{2.2}) into (\ref{2.3}) and integration gives the following linear relationship between $\delta_0$ and $F$,
\begin{equation}\label{2.4}
\delta_0=\frac{2F}{\pi E_*}\,.
\end{equation}
This relationship is rarely listed or discussed in the contact mechanics literature, because it has been well-recognized
that the actual indentation depth is different from the height of the contact zone ($\delta\neq\delta_0$) and that it depends on the boundary conditions. 

\section{Nonlinear $\boldsymbol{\delta}=\boldsymbol{\delta}{\bf (F,b)}$ relationship}

The full elasticity solution of frictionless cylindrical indentation, which specifies the stress and displacement fields in the entire half-space, is available in the literature,
where it is most commonly derived by using complex potentials, e.g., \cite{Bower}, p. 285. From this solution, it follows that the vertical displacement $w=w(x)$ at any point on the surface of a half-space under semi-elliptical pressure distribution, subject to the displacement boundary condition $w(\pm b)=0$, is
\begin{figure}
\centering{
\includegraphics[scale=0.65]{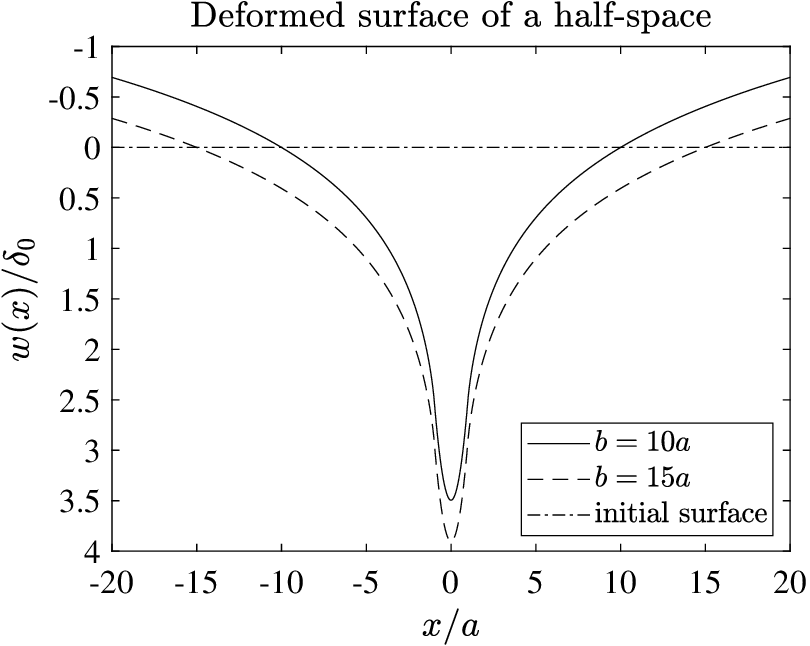}
}
\caption{\label{Fig2} The normalized displacement $w(x)/\delta_0$ vs. $x/a$ for two values of $b$.
The initial position of the free surface of the half-space is
shown by a dash-dotted line. The two shown curves differ by vertical translation of amount $0.4\delta_0$, where $\delta_0=2F/\pi E_*$.
}
\end{figure}
\begin{equation}\label{3.1}
w(x)=\begin{cases}
\delta-\delta_0(x/a)^2\,, \quad x\leq a\,,\\
\delta-\delta_0\left[(x/a)^2-(x/a)\,\sqrt{(x/a)^2-1}+\ln\left(x/a+\sqrt{(x/a)^2-1}\right)\right], \quad x\geq a\,,
\end{cases}
\end{equation}
where
\begin{equation}\label{3.2}
\frac{\delta}{\delta_0}=\beta^2-\beta\,\sqrt{\beta^2-1}+\ln\left(\beta+\sqrt{\beta^2-1}\right),\quad \beta=b/a\,.
\end{equation}
The plots of $w(x)$ vs. $x/a$ from (\ref{3.1}) are shown in Fig. \ref{Fig2}.
Two values of $\beta=b/a$ are used, corresponding to $b=10a$ and $b=15a$,
where $a=(F/F_*)^{1/2}R$ and $F_*=(\pi/4)E_*R$. The displacement $w(x)$ is normalized with $\delta_0=(R/2)F/F_*$. The value $w(0)$ in the case $b=10a$ is $\delta=3.5\delta_0$, while $\delta=3.9\delta_0$ for $b=15a$. The two displacements differ by the vertical translation $\delta_{b=15a}-\delta_{b=10a}=0.4\delta_0$.

For large values of $b/a$, (\ref{3.2}) simplifies to
\begin{equation}\label{3.3}
\delta=\delta_0\left(\ln\frac{2b}{a}+\frac{1}{2}\right),
\end{equation}
which demonstrates the logarithmic increase of $\delta/\delta_0$ with the increase of $b/a$.

\begin{figure}[ht!]
\centering{
\includegraphics[scale=0.65]{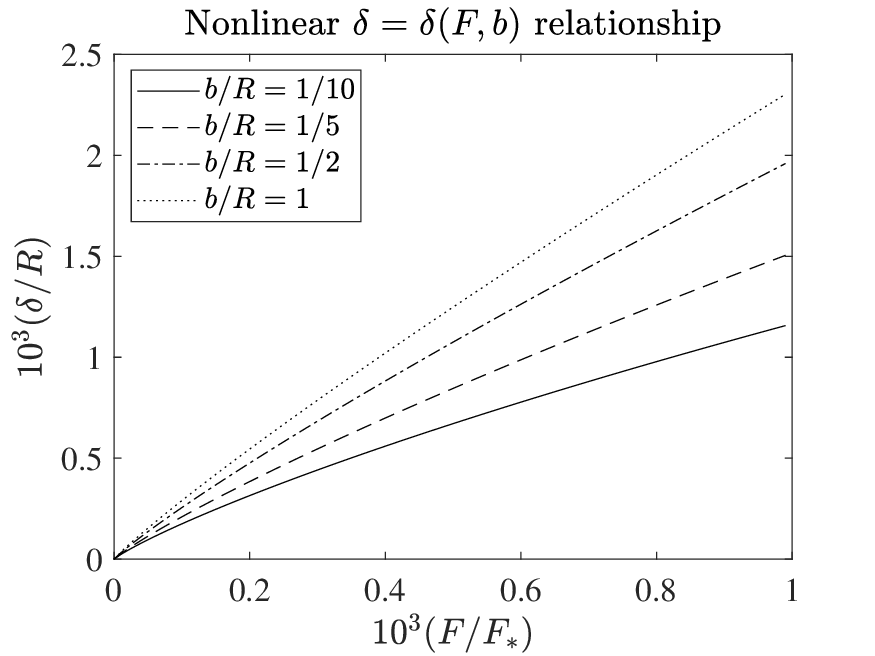}
}
\caption{\label{Fig3} The nonlinear variation of the normalized indentation depth $10^3(\delta/R)$ with the normalized indentation force $10^3(F/F_*)$, where
$F_*=\pi E_*R/4$. The nonlinearity is particularly pronounced for the smaller values of the ratio $b/R$.}
\end{figure}

The nonlinear relationship between $\delta$ and $F$ can be readily recognized from (\ref{3.2}) by observing that
\begin{equation}\label{3.4}
\delta_0=\frac{1}{2}\,Rf\,,\quad a=Rf^{1/2}\,,\quad \beta=\frac{b}{R}\,f^{-1/2}\,,\quad f=\frac{F}{F_*}\,.
\end{equation}
Substituting (\ref{3.4}) into (\ref{3.2}) gives
\begin{equation}\label{3.5}
\delta=\frac{R}{2}\left\{\left(\frac{b}{R}\right)^2-\frac{b}{R}\sqrt{\left(\frac{b}{R}\right)^2-f}+f\ln\left[\frac{b}{R}+\sqrt{\left(\frac{b}{R}\right)^2-f}\,\right]-\frac{1}{2}\,f\ln f\right\}.
\end{equation}
This is an explicit form of the nonlinear force-indentation relationship $\delta=\delta(F,b)$, which is plotted in Fig. \ref{Fig3} for several values of the ratio $b/R$.
The nonlinearity of $\delta=\delta(F,b)$ relationship is more pronounced for smaller values of $b$. The maximum applied force for the indentation to remain elastic is $F_{\rm max}=10^{-n}F_*$, where for metals typically $n\ge 3$.

\subsection{Minimum values ${\bf b_{\rm min}}$ and $\boldsymbol{\delta}_{\rm min}$}

The material of the indented half-space, with a specified displacement boundary condition to eliminate the rigid-body translation, resists the penetration of the indenter. If the points of the free surface $x=\pm b$ are required to have zero vertical displacement throughout the indentation process, as the indentation force increases from $0$ to
its final value $F$, while the contact width increases from $0$ to $2a=2(F/F_*)^{1/2}R$, and the depth of indentation increases from $0$ to its final value $\delta$, there is a minimum value of $b>a$ for which such indentation is geometrically and physically possible. For example, one cannot produce the penetration depth $\delta=\delta_0=2F/(\pi E_*)$, if it is required that $b=a$, i.e., $w(\pm a)=0$, and that the indenter and the free surface of a half-space are both initially at $z=0$. This can be recognized from the fact that the work
(per unit length of the cylinder) of the force $F$ on the
displacement $\delta_0$ is not equal to the work of the pressure $p(x)$ on the vertical displacement $w_a(x)$, i.e.,
\begin{equation}\label{3.6}
\int_0^{\delta_0}\hat{F}(\hat{\delta})\mathrm{d}\hat{\delta}\neq \frac{1}{2}\int_{-a}^ap(x)w_a(x)\mathrm{d}x\,,\quad w_a(x)=\delta_0\left(1-\frac{x^2}{a^2}\right)\,.
\end{equation}
Indeed, the integral on the left-hand side of (\ref{3.6}) is equal to $2W_0/3$ and the integral on the right-hand side is equal to $3W_0/4$, where $W_0=(1/2)F\delta_0=F^2/\pi E_*$ (see below). This means that one needs to find the minimum value of $b>a$, and the corresponding value of $\delta>\delta_0$, such that
\begin{equation}\label{3.7}
W_F=\int_0^{\delta}\hat{F}(\hat{\delta})\mathrm{d}\hat{\delta}\equiv W_p=\frac{1}{2}\int_{-a}^ap(x)w_b(x)\mathrm{d}x\,,\quad w_b(x)=\delta-\delta_0\,\frac{x^2}{a^2}\,.
\end{equation}
The indentation work condition $W_F=W_p$ must hold because the indentation stress and strain fields in the half-space are produced by the indentation force $F$, transmitted to the half-space by the statically equivalent pressure distribution $p(x)=(2F/\pi a)(1-x^2/a^2)^{1/2}$. For example, for spherical indentation of a half-space $F\sim \delta^{3/2}$ and $W_F=W_p=(2/5)F\delta$, while for conical indentation $F\sim \delta^2$ and $W_F=W_p=(1/3)F\delta$.

The expressions for the works appearing in (\ref{3.7}) can be conveniently determined from
\begin{equation}\label{3.8}
W_F=\int_0^\delta \hat{F}(\hat{\delta})\mathrm{d}\hat{\delta}=F\delta-\int_0^F \hat{\delta}\mathrm{d}\hat{F}\,,
\end{equation}
\begin{equation}\label{3.9}
W_p=\frac{1}{2}\int_{-a}^ap(x)w_b(x)\mathrm{d}x=\frac{1}{2}\,F\delta-\frac{1}{8}\,F\delta_0\,.
\end{equation}
Substituting expression (\ref{3.2}) for $\delta$, the pressure work $W_p$ in (\ref{3.9}) becomes
\begin{equation}\label{3.10}
W_p=W_0\left[\beta^2-\beta\sqrt{\beta^2-1}+\ln\left(\beta+\sqrt{\beta^2-1}\right)-\frac{1}{4}\right],\quad W_0=\frac{1}{2}\,F\delta_0=\frac{F^2}{\pi E_*}\,.
\end{equation}
In particular, if $\beta=1$ this gives $W_p=3W_0/4$.

The complementary work $\int_0^F\hat{\delta}\,\mathrm{d}\hat{F}$ of the force $F$, appearing in the expression for $W_F$ in (\ref{3.8}), can also be evaluated analytically. Toward that, it is convenient to rewrite the expression for $\hat{\delta}$ from (\ref{3.2}) as
\begin{equation}\label{3.11}
\hat{\delta}=\frac{1}{2R}\!\left(b^2-b\sqrt{b^2-\hat{a}^2}+\hat{a}^2\ln\frac{b+\sqrt{b^2-\hat{a}^2}}{\hat{a}}\right),
\end{equation}
and conveniently use $\hat{a}^2$ as the integration variable, such that $\mathrm{d}\hat{F}=(\pi E_*/4R)\,\mathrm{d}(\hat{a}^2)$. Upon a somewhat lengthy integration, it follows that
\begin{figure}
\centering{
\includegraphics[scale=0.5]{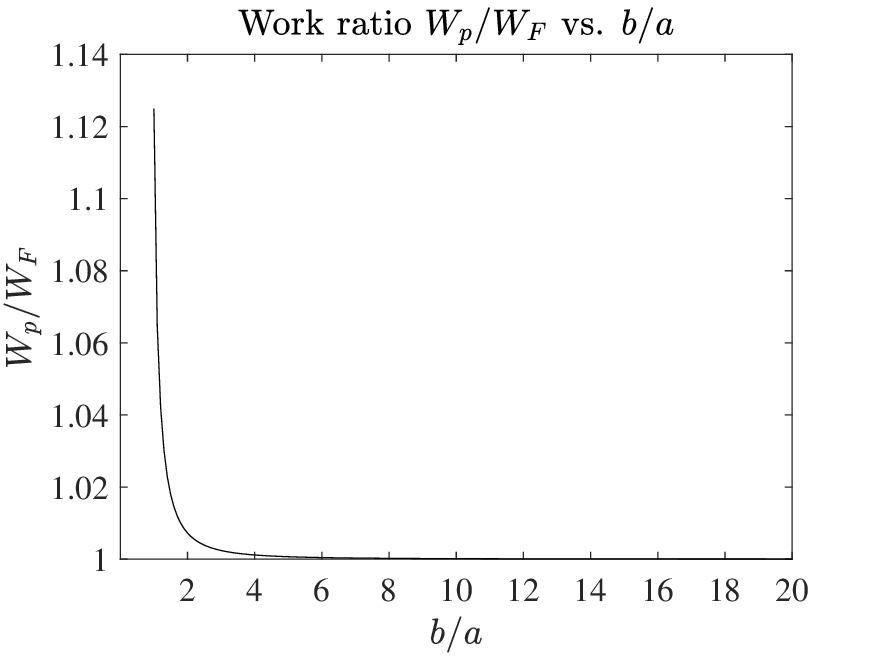}
\hskip5mm
\includegraphics[scale=0.5]{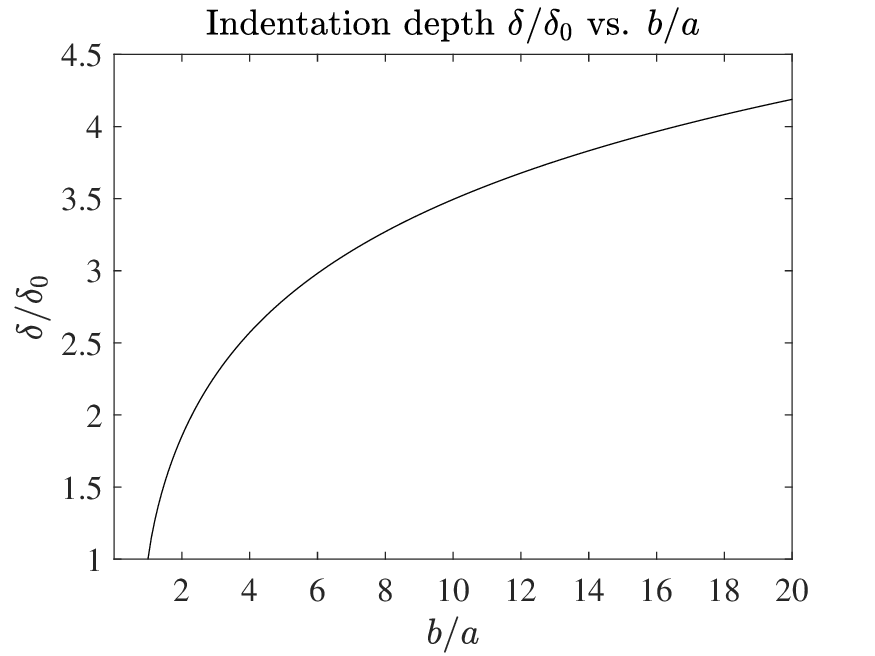}
}
\vskip3mm
\centerline{(a)\hskip75mm (b)}
\caption{\label{Fig4} (a) The decrease of the work ratio $W_p/W_F$ toward the value of 1 with the increase of $b/a$. For $b>b_{\rm min}\approx 10a$, the work ratio falls below the value of $1.00013$. (b) The variation of $\delta/\delta_0$ with $b/a$. The values of $\delta$ for $b\ge b_{\min}\approx 10a$ correspond to actual indentation.}
\end{figure}
\begin{equation}\label{3.12}
W_F=W_0\left[\frac{2}{3}\,\beta^4-\frac{1}{3}\,\beta (1+2\beta^2)\sqrt{\beta^2-1}+\ln\left(\beta+\sqrt{\beta^2-1}\right)\right].
\end{equation}
If $\beta=1$, this gives $W_F=2W_0/3$, which is different from $W_p=3W_0/4$, their ratio being $W_p/W_F=9/8=1.125$.

The plot of the work ratio $W_p/W_F$ versus $b/a$ is shown in Fig. \ref{Fig4}a. For $\beta=10$, (\ref{3.10}) and (\ref{3.12}) give
$W_p=3.244479W_0$ and $W_F=3.244061W_0$, which can be considered to be approximately equal to each other,
the relative error being $(W_p-W_F)/W_p=1.3\times 10^{-4}W_0$. We therefore define $b_{\rm min}\approx 10a$, the corresponding (minimum) indentation depth being  $\delta_{\rm min}\approx 3.5\delta_0$. For $\beta\gg 1$, (\ref{3.10}) and (\ref{3.12}) both become
\begin{equation}\label{3.13}
W_p=W_F=W_0\left(\frac{1}{4}+\ln\frac{2b}{a}\right),
\end{equation}
which corresponds to the horizontal plateau $W_p/W_F=1$ in Fig. \ref{Fig4}a, rapidly approached as $b/a$ is increased.

The variation of $\delta$ with $b/a$ is shown in Fig. \ref{Fig4}b for the values of $b\ge a$, keeping in mind that the actual indentations correspond to $b\ge b_{\rm min}$. For large values of $\beta=b/a$, the approximation applies
\begin{equation}\label{3.14}
\delta=\delta_0(k+\ln\beta)\,,\quad k=\frac{1}{2}+\ln 2 \approx 1.1931\,.
\end{equation}
For example, the indentation depth is $\delta\approx 8.1\delta_0$ for $b=10^3a$, while $\delta\approx 10.4\delta_0$ for $b=10^4a$. From (\ref{3.14}), one can also express explicitly the
value of $b=a\exp(\delta/\delta_0-k)$ corresponding to any given indentation depth $\delta\ge \delta_{\rm min}\approx 3.5\delta_0$. Although all values of $b\ge b_{\rm min}$ are conceptually important, large values of $b$, much greater than $b_{\rm min}$, are most useful for the analysis of indentation of large, but finite size blocks of material.
%

\section{Displacement along vertical axis below the indenter}

The vertical displacement $u_z$ along the $z$ axis beneath the indenter can be obtained from the complete solution of the cylindrical indentation problem by integrating the longitudinal strain $\epsilon_z=\partial u_z/\partial z$ along the $z$ axis, and by specifying the integration constant from the condition $u_z(z=0)=\delta$. The longitudinal strain is, by Hooke's law,
\begin{equation}\label{4.1}
\epsilon_z=\frac{1+\nu}{E}\,[(1-\nu)\sigma_z-\nu\sigma_x]\,,
\end{equation}
where the stresses along the $z$ axis are (e.g., \cite{Fischer}, p. 92)
\begin{equation}\label{4.2}
\sigma_x=-\frac{2F}{\pi a}\left[(1+2\zeta^2)(1+\zeta^2)^{-1/2}-2\zeta\right],\quad \sigma_z=-\frac{2F}{\pi a}\,(1+\zeta^2)^{-1/2}\,,\quad \zeta=\frac{z}{a}\,.
\end{equation}
Thus, upon integration, the displacement along the $z$ axis is found to be
\begin{figure}
\centering{
\includegraphics[scale=0.65]{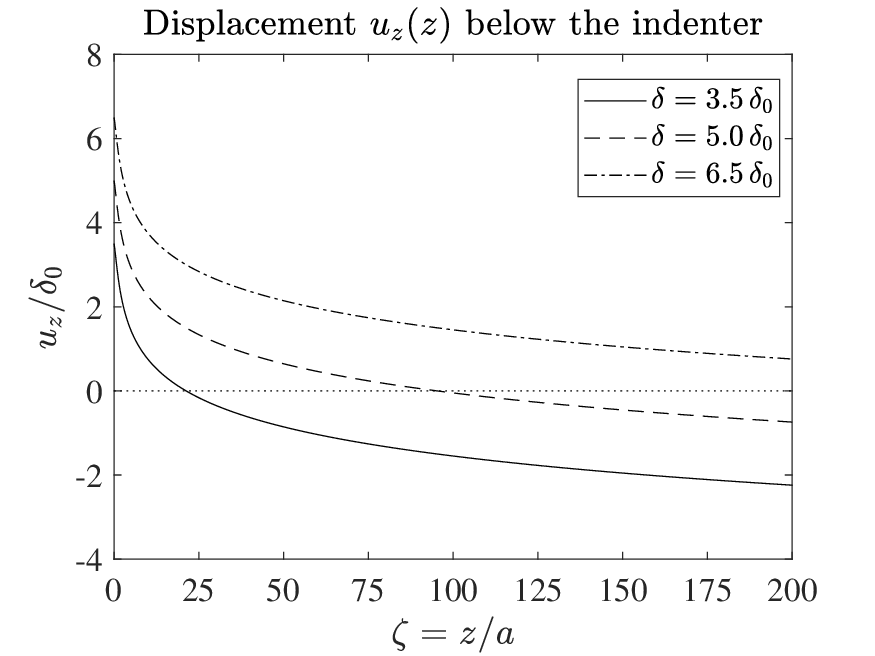}
}
\caption{\label{Fig5} The variation of the normalized displacement $u_z(z)/\delta_0$ with $\zeta=z/a$ for $\nu=1/3$ and for three shown values of the indentation depth $\delta$, where $\delta_0=2F/\pi E_*$ and $a=(4RF/\pi E_*)^{1/2}$. The selected values of $\delta$ correspond to the $h$-datums of $21a$, $95a$ and $427a$, as discussed in the text.
}
\end{figure}
\begin{equation}\label{4.3}
u_z(z)=\delta-\delta_0\left[\ln\left(\zeta+\sqrt{1+\zeta^2}\right)+\frac{\nu}{1-\nu}\,\zeta\left(\zeta-\sqrt{1+\zeta^2}\right)\right].
\end{equation}
The plots of $u_z(z)/\delta_0$ vs. $\zeta=z/a$ are shown for three selected values of the indentation depth $\delta$ in Fig. \ref{Fig5}. The displacement $u_z$ beneath the indenter is at first downwards (for sufficiently small $z$), but then it turns its direction upwards. For $z\gg a$, (\ref{4.3}) simplifies to
\begin{equation}\label{4.4}
u_z(z)=\delta-\delta_0\left[\ln\frac{2z}{a}-\frac{\nu}{2(1-\nu)}\right],\quad z\gg a\,.
\end{equation}

Expression (\ref{4.4}) can be used in an approximate analysis of the indentation of a large elastic block bonded at its bottom to a rigid substrate.
Suppose that, for a given indentation force $F$, one wants to specify the value of $h$, and thus $\delta$, such that $u_z(z=h\gg a)=0$. From (\ref{4.4}), it then follows that
\begin{equation}\label{4.5}
\delta=\delta_0\left[\ln\frac{2h}{a}-\frac{\nu}{2(1-\nu)}\right].
\end{equation}
If $\delta_0$ in (\ref{4.5}) is replaced with $2F/(\pi E_*)$, as given by (\ref{2.4}), expression (\ref{4.5})
coincides with expression (5.58), p. 131, of \cite{Johnson}, which was used in \cite{Barber}, p. 101, to discuss the indentation of a large elastic block bonded at its bottom to a rigid substrate.

The relationship between the height $h$ and the horizontal distance $b$ used to define the displacement datum $x=\pm b$ on the free surface in section 3 can also be established.
When (\ref{4.5}) is equated to (\ref{3.3}), if follows
\begin{equation}\label{4.6}
\ln\frac{2h}{a}-\frac{\nu}{2(1-\nu)}=\frac{1}{2}+\ln\frac{2b}{a}\,.
\end{equation}
This gives the linear relationship between $b$ and $h$,
\begin{equation}\label{4.7}
b=h\exp\left[-\frac{1}{2(1-\nu)}\right],
\end{equation}
with the coefficient dependent on the Poisson ratio only.
For example, if $h=10^3a$ then $b=472.4a$, with the corresponding depth of indentation $\delta=7.35\delta_0$; if $h=10^4a$ then $b=4724a$ and $\delta=9.65\delta_0$. In both cases, the Poisson ratio was taken to be $\nu=1/3$. Although derived for large values of $b$ and $h$, the linear relationship (\ref{4.7}) approximately holds even for the values 
of $b$ and $h$ near $b_{\rm min}$ and $h_{\rm min}$, as recognized by numerical evaluations (see Fig. \ref{Fig8} below).

\section{Nonlinear $\boldsymbol{\delta}=\boldsymbol{\delta}{\bf (F,h)}$ relationship}

The nonlinear relationship $\delta=\delta(F,h)$ can be derived from (\ref{4.3}). If the displacement is required to be zero at the point $z=h$ below the indenter, from (\ref{4.3})
it is found that
\begin{equation}\label{5.1}
\delta=\delta_0\left[\ln\left(\eta+\sqrt{1+\eta^2}\right)+\frac{\nu}{1-\nu}\,\eta\left(\eta-\sqrt{1+\eta^2}\right)\right],\quad \eta=\frac{h}{a}\,.
\end{equation}
Because $\delta_0=(F/F_*)R/2$ and $a=(F/F_*)^{1/2}R$,  expression (\ref{5.1}) can be rewritten as
\begin{equation}\label{5.2}
\frac{\delta}{R}=\frac{1}{2}\,f\!\left[\ln\left(\eta/\sqrt{f}+\sqrt{1+\eta^2/f}\,\right)+\frac{\nu}{1-\nu}\,(\eta/\sqrt{f})\left(\eta/\sqrt{f}-\sqrt{1+\eta^2/f}\,\right)\right],\quad f=\frac{F}{F_*}\,.
\end{equation}
This is an explicit form of the nonlinear relationship $\delta=\delta(F,h)$. The plot of $\delta/\delta_0$ vs. $h/a$ is shown in Fig. \ref{Fig6}a. For $\eta^2/f=(h/R)^2(F_*/F)\gg 1$, (\ref{5.2}) simplifies to
\begin{equation}\label{5.3}
\frac{\delta}{R}=\frac{1}{4}\,\frac{F}{F_*}\!\left(\ln\frac{F_*}{F}+2\ln\frac{2h}{R}-\frac{\nu}{1-\nu}\right).
\end{equation}
The relationship (\ref{5.2}) for $\delta=\delta(F,h)$ involves the Poisson ratio $\nu$, while the relationship (\ref{3.5}) for $\delta=\delta(F,b)$ does not (apart from the dependence of $F_*$ on $\nu$), which is as expected because the relationship between $h$ and $b$, such as (\ref{4.7}), involves the Poisson ratio.

\subsection{Minimum values ${\bf h_{\rm min}}$ and $\boldsymbol{\delta}_{\rm min}$}

Similarly to the analysis of $b_{\rm min}$ and the corresponding $\delta_{\rm min}$ from section 3, one can determine $h_{\rm min}$ and the corresponding $\delta_{\rm min}$
by requiring that $W_F=W_p$ (to within a prescribed numerical accuracy). By the same analysis as in section 3, it follows that the pressure work is
\begin{equation}\label{5.4}
W_p=W_0\left[\frac{\nu}{1-\nu}\left(\eta^2-\eta\sqrt{1+\eta^2}\right)+\ln\left(\eta+\sqrt{1+\eta^2}\right)-\frac{1}{4}\right],\quad W_0=\frac{1}{2}\,F\delta_0=\frac{F^2}{\pi E_*}\,.
\end{equation}
while the work of the force $F$ is
\begin{figure}
\centering{
\includegraphics[scale=0.5]{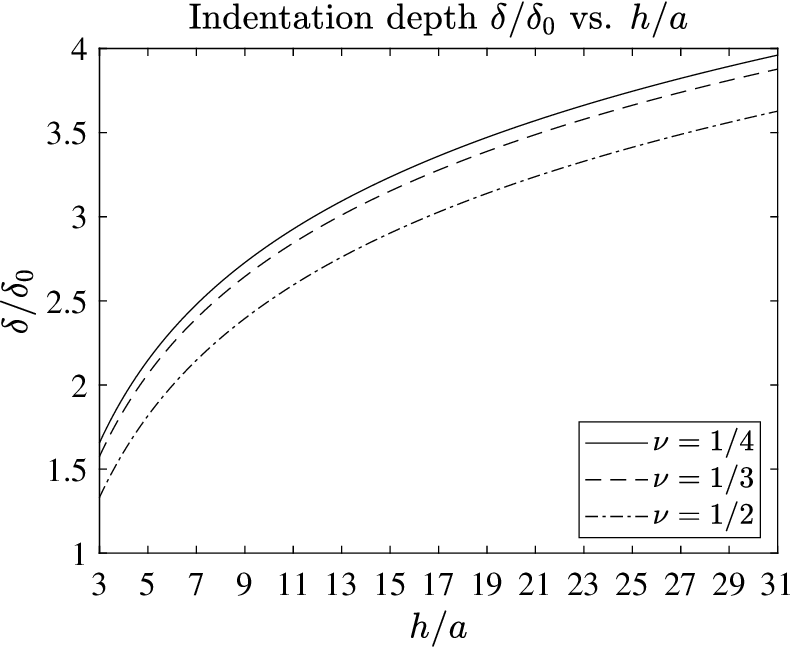}\hskip5mm
\includegraphics[scale=0.5]{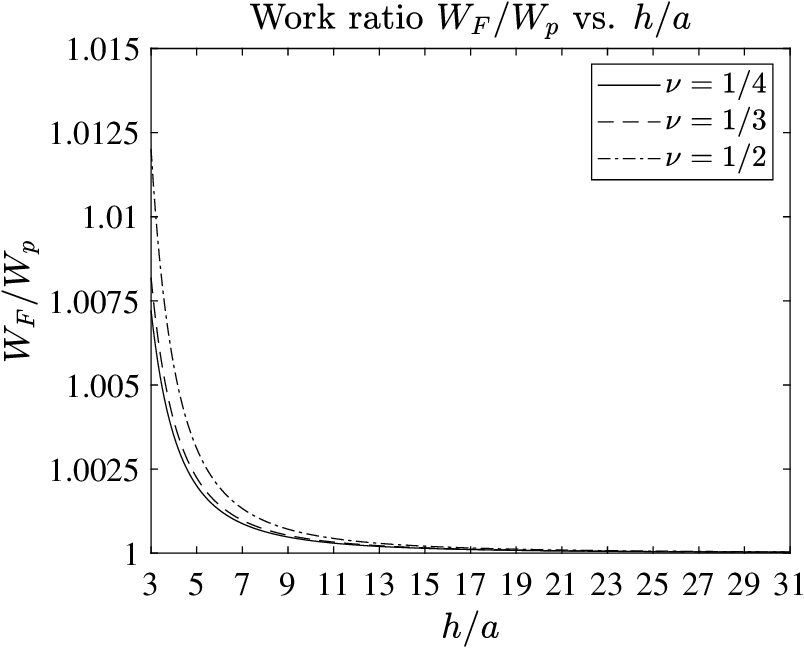}
}
\vskip3mm
\centerline{(a)\hskip75mm (b)}
\caption{\label{Fig6} (a) The variation of $\delta/\delta_0$ with $h/a$. The results are relevant for $h\ge h_{\min}$, dependent on the Poisson ratio, as discussed in the body of the paper. For large values of $h/a$,
the normalized indentation depth is approximately $\delta/\delta_0=\ln(2h/a)-\nu/[2(1-\nu)]$. (b) The decrease of the work ratio $W_F/W_p$ toward the value of 1 with the increase of $h/a$. Note that the normalizing factors $\delta_0$ and $a$ are also dependent on the Poisson ratio, because $\delta_0=(1-\nu^2)(2F/\pi E)$ and $a=[4(1-\nu^2)FR/\pi E]^{1/2}$.
}
\end{figure}
\begin{equation}\label{5.5}
\begin{split}
W_F&=W_0\Big\{\frac{2\nu}{1-\nu}\left[\frac{2}{3}\,\eta (1+\eta^2)^{3/2}-\eta\sqrt{1+\eta^2}-\frac{2}{3}\,\eta^4\right]+\ln\left(\eta+\sqrt{1+\eta^2}\right)\\
&-\frac{2}{3}\,\eta^4-\frac{1}{3}\,\eta(1-2\eta^2)\sqrt{1+\eta^2}\,\Big\}\,.
\end{split}
\end{equation}
The plot of the work ratio $W_F/W_p$ with $\eta=h/a$ is shown in Fig. \ref{Fig6}b for three selected values of the Poisson ratio. For $\eta\gg 1$, the work expressions (\ref{5.4}) and (\ref{5.5}) both simplify to
\begin{equation}\label{5.6}
W_p=W_F=W_0\left[-\frac{1+\nu}{4(1-\nu)}+\ln\frac{2h}{a}\right],
\end{equation}
which explains the horizontal plateau $W_F/W_p=1$ in Fig. \ref{Fig6}b, rapidly approached with the increase of $h/a$. For $h=a$, the two works are clearly different,
\begin{equation}\label{5.7}
W_p=W_0\left[\ln (1+\sqrt{2})-\frac{\nu}{1-\nu}\left(\sqrt{2}-1\right)-\frac{1}{4}\right],\quad
W_F=W_0\left[\ln (1+\sqrt{2})-\frac{1+\nu}{1-\nu}\,\frac{2-\sqrt{2}}{3}\right].
\end{equation}
\begin{figure}
\centering{
\includegraphics[scale=0.65]{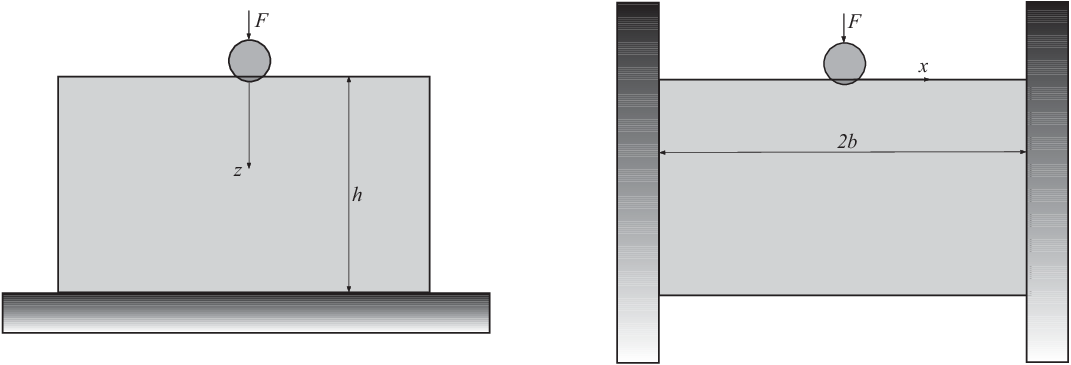}
}
\vskip3mm
\centerline{(a)\hskip65mm (b)}
\caption{\label{Fig7}  (a) Indentation of a large elastic block supported by a rigid substrate at its bottom $z=h$, or (b) laterally at its sides $x=\pm b$.
}
\end{figure}
For example, for $\nu=1/3$, (\ref{5.7}) gives $W_p=0.4243W_0$ and $W_F=0.4908W_0$, their ratio being $W_F/W_p=1.157$. This means that one cannot produce indentation with the contact width $2a$ if it is required that $u_z(z=a)=0$ and that the indenter and the free surface of a half-space are both initially at $z=0$. As a matter of fact, if $h=a$, the indentation depth predicted by (\ref{5.1}) would be less than $\delta_0$, i.e., $\delta\approx 0.674\delta_0$, meaning that in section 3 $b<a$, which does not correspond to physical indentation from the $z=0$ initial surface level. 

The minimum indentation depth $\delta_{\rm min}\approx 3.5\delta_0=3.5(1-\nu^2)(2F/\pi E)$ obtained in section 3 when $b=b_{\rm min}\approx 10\delta_0=10(1-\nu^2)(2F/\pi E)$, corresponds to $h_{\rm min}=16.45a$ in the case $\nu=0$, $h_{\rm min}=19.44a$ in the case $\nu=1/4$, $h_{\rm min}=21.13a$ in the case $\nu=1/3$, and $h_{\rm min}=27.13a$ in the case $\nu=1/2$, where $a=[4(1-\nu^2)FR/\pi E]^{1/2}$.

As discussed earlier, the $h$-datum is of practical interest for an approximate analysis of the indentation of a thick block constrained at its bottom by being bonded to a rigid substrate
(Fig. \ref{Fig7}a). In this case the values of $b$ corresponding to $h\gg a$, define the portion of the free surface $|x|<b$ which is displaced downwards relative to the initial un-indented free surface $z=0$, while the points $|x|>b$ are displaced upwards. Furthermore, the $b$-datum itself, with a large value of $b\gg a$, may be of interest for an approximate analysis of the indentation of a wide block of material supported laterally, rather than at its bottom, in which case the constraints at $x=\pm b$ prevent vertical displacement of these points (Fig. \ref{Fig7}b). In both cases shown in Fig. \ref{Fig7} the stress and strain fields in the contact region are nearly equal to that in a half-space, and the dominant term in the expression for the indentation depth $\delta/\delta_0$ is still the logarithmic term  $\ln(2b/a)$ or $\ln(2h/a)$; see (\ref{3.3}) and (\ref{4.5}). This can be verified by using the finite element method to satisfy the remote displacement and traction boundary conditions at $z=h$ and $x=\pm b$, which is beyond the scope of the present paper.

The general relationship between $b$ and $h$, corresponding to the same indentation depth $\delta$, is obtained by equating expressions (\ref{3.2}) and (\ref{5.1}),
\begin{figure}
\centering{
\includegraphics[scale=0.5]{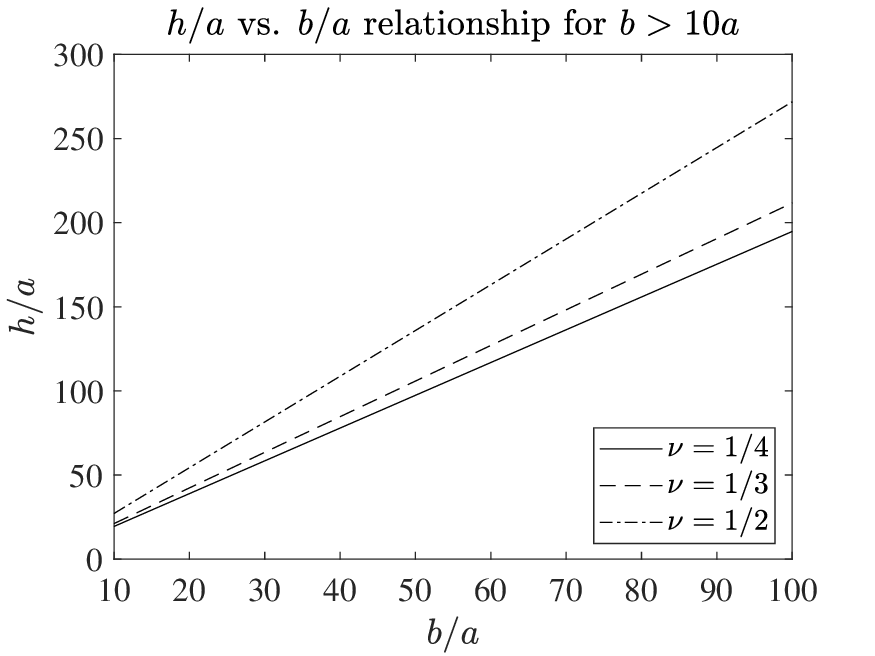}\hskip5mm
\includegraphics[scale=0.5]{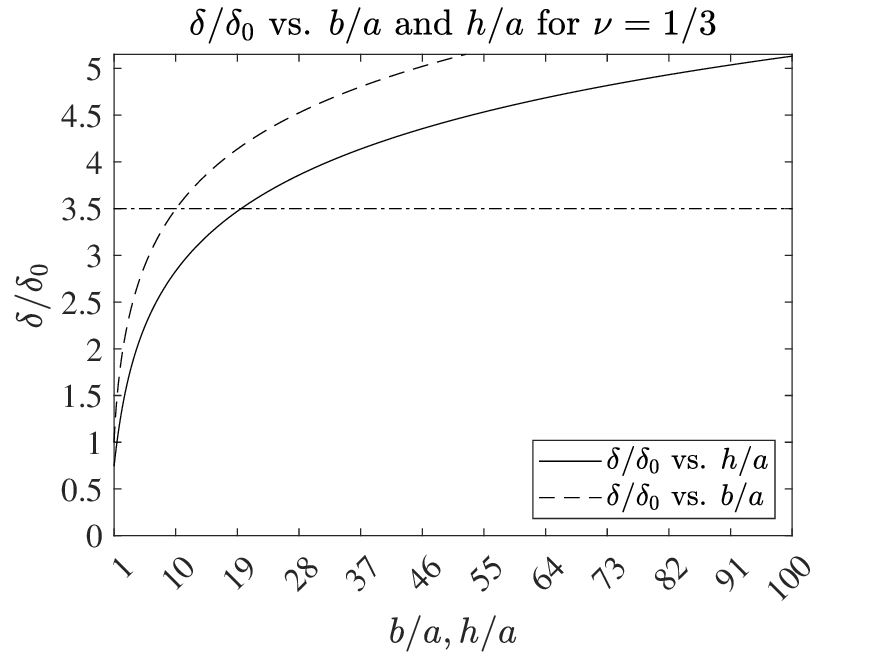}\hskip5mm
}
\caption{\label{Fig8}  (a) The relationship between $h/a$ and $b/a$, corresponding to the same indentation $\delta$, as obtained from (\ref{5.7}). For large values of $b/a$,
the linear relationship (\ref{4.7}) holds, with the slope $h/b=\exp[0.5/(1-\nu)]$. (b) The variation of $\delta/\delta_0$ vs. $b/a$ and $h/a$ in the case $\nu=1/3$. The actual indentation depths are above the minimum indentation depth $\delta_{\rm min}\approx 3.5\delta_0$, shown by the dotted horizontal line.
}
\end{figure}
\begin{equation}\label{5.8}
\beta^2-\beta\,\sqrt{\beta^2-1}+\ln\left(\beta+\sqrt{\beta^2-1}\right)=\ln\left(\eta+\sqrt{1+\eta^2}\right)+\frac{\nu}{1-\nu}\,\eta\left(\eta-\sqrt{1+\eta^2}\right).
\end{equation}
If this is numerically solved for $\eta=h/a$ in terms of $\beta=b/a$, the plots shown in Fig. \ref{Fig8}a are obtained. The relationship between $h$ and $b$ is almost linear in the entire range of $b/a>50$ or so, particularly for greater values of $\nu$. For large values of $b/a$, it is well approximated by  $h=b\exp[0.5/(1-\nu)]$, following from (\ref{4.7}). Figure \ref{Fig8}b shows the variation of $\delta/\delta_0$ vs. $b/a$ and $h/a$ in the case $\nu=1/3$, as determined from (\ref{3.2}) and (\ref{5.1}). The actual indentation depths are above the minimum indentation depth $\delta_{\rm min}=3.5\delta_0$, shown by a dash-dotted horizontal line.

\section{Conclusion}

A closed-form nonlinear relationship $\delta=\delta(F,b)$ between the depth of indentation $\delta$ and the applied force $F$ for cylindrical indentation of an elastic half-space, with the displacement datum at the points $x=\pm b$ of the free surface, is derived and is given by (\ref{3.2}).
The requirement that the work done by the indentation force must be equal to the work done by the contact pressure ($W_F=W_p$) specifies the
minimum value of $b$ for which the indentation is geometrically and physically possible, and thus the minimum value of the indentation depth $\delta$. The value of $b_{\rm min}$ is found to be about 10 times greater than the semi-width of the contact zone
$a=(4RF/\pi E_*)^{1/2}$, where $R$ is the radius of the cylindrical indenter and $E_*=E/(1-\nu^2)$ is the effective modulus of elasticity of the indented half-space.
The corresponding minimum indentation depth is
$\delta_{\rm min}\approx 3.5\delta_0$, where $\delta_0=2F/\pi E_*$ is the height of the contact zone, which is independent of $b$.
When the displacement datum is taken to be at a point along the axis of symmetry below the load, at some distance $h$ from the free surface, there is a closed-form expression for $\delta=\delta(F,h)$, which is dependent on the Poisson ratio and is given by (\ref{5.2}). The work requirement $W_F=W_p$ specifies again the
minimum value of $h$ for which the indentation is geometrically and physically possible. This value is found to be
$h_{\rm min}\approx 16.5a$ in the case $\nu=0$, about $21a$ in the case $\nu=1/3$, and about $27a$ in the case $\nu=1/2$,
with the corresponding minimum indentation depth $\delta_{\rm min}\approx 3.5\delta_0$. The relationship between $h$ and $b$, corresponding to the same indentation depth $\delta\ge \delta_{\rm min}$, is shown to be linear for large values of $h$ and $b$. The obtained results may be useful in the analysis of the indentation of elastic blocks of finite size bonded to a rigid substrate laterally or at its bottom.

%


\noindent{\bf ORCID IDs}\\
\noindent Vlado A. Lubarda 0000-0002-0474-6681\\
\noindent Marko V. Lubarda 0000-0002-3755-271X

\vskip2mm
\noindent {\bf Author contribution statement:} The authors contributed equally to the contents of the paper.
\vskip2mm
\noindent{\bf Disclosure statement:} The authors report there are no competing interests to declare.

\end{document}